\pdfoutput=1

\documentclass[11pt]{article}

\usepackage{acl}

\usepackage{times}
\usepackage{latexsym}

\usepackage[T1]{fontenc}

\usepackage[utf8]{inputenc}

\usepackage{microtype}
\usepackage{graphicx}

\title{Semantic Equivalence of e-Commerce Queries}

\author{Aritra Mandal \and Daniel Tunkelang \and Zhe Wu \\
Ebay Inc. \\
San Jose, California, United States\\
\texttt{arimandal@ebay.com, dtunkelang@ebay.com, zwu1@ebay.com}
} 

\begin{document}
\maketitle
\begin{abstract}

Search query variation poses a challenge in e-commerce search, as equivalent search intents can be expressed through different queries with surface-level differences. This paper introduces a framework to recognize and leverage query equivalence to enhance searcher and business outcomes. The proposed approach addresses three key problems: mapping queries to vector representations of search intent, identifying nearest neighbor queries expressing equivalent or similar intent, and optimizing for user or business objectives. The framework utilizes both surface similarity and behavioral similarity to determine query equivalence. Surface similarity involves canonicalizing queries based on word inflection, word order, compounding, and noise words. Behavioral similarity leverages historical search behavior to generate vector representations of query intent. An offline process is used to train a sentence similarity model, while an online nearest neighbor approach supports processing of unseen queries. Experimental evaluations demonstrate the effectiveness of the proposed approach, outperforming popular sentence transformer models and achieving a Pearson correlation of 0.85 for query similarity. The results highlight the potential of leveraging historical behavior data and training models to recognize and utilize query equivalence in e-commerce search, leading to improved user experiences and business outcomes. Further advancements and benchmark datasets are encouraged to facilitate the development of solutions for this critical problem in the e-commerce domain.

\end{abstract}

\section{Introduction}

In e-commerce search, there can be multiple ways for searchers to express a given search intent. Two queries that express equivalent intent may vary superficially, differing only in word inflection, word order, compounding, or the inclusion of noise words (e.g., "men's backpack" = "back packs for men"). Sometimes, however,  the variation between equivalent queries is not immediately evident from surface query features (e.g., "iphone headphone jack adapter" = "lightning to 3.5mm").

Recognizing when two queries express equivalent intent makes it possible to improve searcher and business outcomes in various ways. For example, if the search engine retrieves different results for the two queries or ranks results in different order, then the search engine can choose the retrieval and ranking strategy that leads to the better user or business outcome. The search engine can implement this optimization by mapping the lower-performing query to the higher-performing query that expresses equivalent intent.

There is also value in recognizing when two queries express similar but not quite equivalent intent. The search engine can leverage this similarity to increase recall while mostly preserving precision. The increase in recall can lead to better user or business outcomes, particularly for queries that would otherwise return no or few results. For example, if the query "14th century music" returns few results, the search engine could also include results for "medieval music".

In order to recognize and leverage query equivalence, we need to solve three problems. First, we need to map a search query to a vector representation of the underlying search intent. Second, we need to identify the nearest neighbors of a query, i.e., other queries that express equivalent or similar intent. Third, we need to choose among these candidates to optimize for a user or business outcome, such as increasing recall or conversion.

In contrast to related work on e-commerce query similarity that relies on co-purchased products to generate training data \cite{Huang2023}, our approach relies on vector representations of products, so that engagement with similar but distinct products still leads to similar query representations.

\section{Surface Similarity}

Sometimes the similarity between two queries is evident from their surface features. We can recognize surface variations among equivalent queries by mapping queries to canonical forms:

\begin{itemize}

\item Word inflection.  Equivalent queries often vary in word inflection. The most common case of variation in word inflection is between the singular and plural forms of nouns (e.g., "birkenstock" = "birkenstocks"). We can canonicalize queries through stemming or lemmatization.

\item Word order. Equivalent queries often use the same words in different order (e.g., "earbuds samsung" = "samsung earbuds"). We can canonicalize queries by sorting the query tokens in alphabetical (lexicographic) order.

\item Compounding. Equivalent queries may vary in compounding or decompounding query tokens (e.g., "backpack" = "back pack"). Variations in compounding are more common in certain languages, such as German. We can canonicalize queries by removing spaces and other separators, such as punctuation.

\item Noise words. Equivalent queries may vary in the inclusion of noise words (e.g., "games for nintendo switch" = "games nintendo switch"). We can canonicalize queries by removing noise words.

\end{itemize}

While surface similarity is highly predictive of query equivalence, there are superficially similar queries that do not express similar, let alone equivalent, intent. Some examples of dissimilar intents despite surface similarity:

\begin{itemize}

\item Word inflection: "blackberry" != "blackberries" -- the singular is strongly associated with once popular brand of smartphones while the plural is mainly associated with the fruit.

\item Word order: "shirt dress" != "dress shirt" -- the first is a type of dress, while the second is a type of shirt.

\item Compounding: "wildcats" != wild cats -- the first refers to a variety of sports teams, while the second refers to animals.

\item Noise words: "desk with lamp" != "desk lamp" -- the first is a desk that includes a lamp, while the second is a type of lamp.

\end{itemize}

Nonetheless, surface similarity is a strong signal of query equivalence. Combining it with a guard rail, such as the observed or predicted result category based on search demand, allows us to establish query equivalence with high confidence.

\section{Behavioral Similarity}

We can look beyond surface similarity and represent the intent of a query as summary of how searchers historically engage with results after performing that query. Since product titles are self-contained and generally longer than queries, at least in the authors' experience working with product titles and queries from large e-commerce platforms such as eBay, we can generate robust embedding vectors from them using freely available models, such as fastText \cite{bojanowski2017enriching}, BERT \cite{devlin2019bert}, or SBERT \cite{reimers2019sentencebert}. We can then obtain the vector representation of a query representation by taking the mean, or some other aggregation, of the embeddings of products that searchers engage with (e.g., click on) after performing that query.

This approach requires that we have a reasonable amount of historical behavior data for a query to aggregate, e.g., at least 20 clicks, in order to obtain a robust query representation. For every query that satisfies this requirement, we can obtain a vector representation of its intent by taking the mean of vectors of its associated product vectors.

We can then compute the similarity between two queries as the cosine similarity of their two vectors. We have observed, again in our experience working with queries from large e-commerce platforms such as eBay, that queries that express equivalent or near-equivalent intent have a very high cosine similarity, generally above 0.98.

\section{Using Offline Pairs for Training}

Unfortunately, this approach's requirement of historical behavior data, as well as its use of offline computation, restricts its applicability to relatively frequent queries, typically those that appear in the head and torso of the query distribution. Addressing the long tail of rare queries, including previously unseen queries, requires a different approach that computes query vectors online rather than relying on offline aggregation.

To implement an online approach, we can train a sentence similarity model that allows us to directly transform two queries into vectors and obtain a similarity score from their cosine similarity. We train this model using pairs of queries obtained from the previously described offline process. We can then create a collection of (query, query, similarity) triples, and use this collection to train an online sentence transformer model for query similarity.

This approach allows us to generalize our offline behavioral model to an online setting, but it encounters two challenges.

The first challenge is choosing which offline query pairs to use for training. If we sample them using a uniform random process, most of the cosine similarities will be much less than 1. Our model might be able to distinguish a query pair with a cosine similarity of 0.4 from a query pair with a cosine similarity of 0.6 -- but that is not very helpful for determining whether two query express equivalent or near-equivalent intent. We want to train a model that is sensitive in the range that matters -- namely for query pairs whose cosine similarity is close to 1. In order to do so, we oversample query pairs with cosines close to 1. We obtain this sample by populating a nearest neighbor database (we used FAISS \cite{johnson2017billionscale}) with the query vectors.

The second challenge is ensuring that we represent enough signal in the vectors to cover the heterogeneity of query intents, especially when they are spread across large variety of product categories. In particular, our collection of query pairs may not teach a model how to distinguish queries that have high surface similarity but are associated with different product categories (e.g., "blackberry" and "blackberries"). We can address this challenge by using a separately trained query category classifier that maps queries to categories, specifically a fastText model trained from historical click data. Building such a classifier is far simpler than building a model for query similarity, since it operates at a much coarser granularity of the query intent space. We can then include the output of this classifier as an input for the query similarity model. In order to take advantage of the rich semantic signal in category names and taxonomy paths, we take this verbose representation, e.g., representing the category of "mens levis 501" as "Clothing, Shoes \& Accessories -> Men -> Men's Clothing -> Men's Jeans," and concatenate it to the query before applying sentence transformer model to the resulting string.

\begin{figure}[t]
\includegraphics[width=6.5cm]{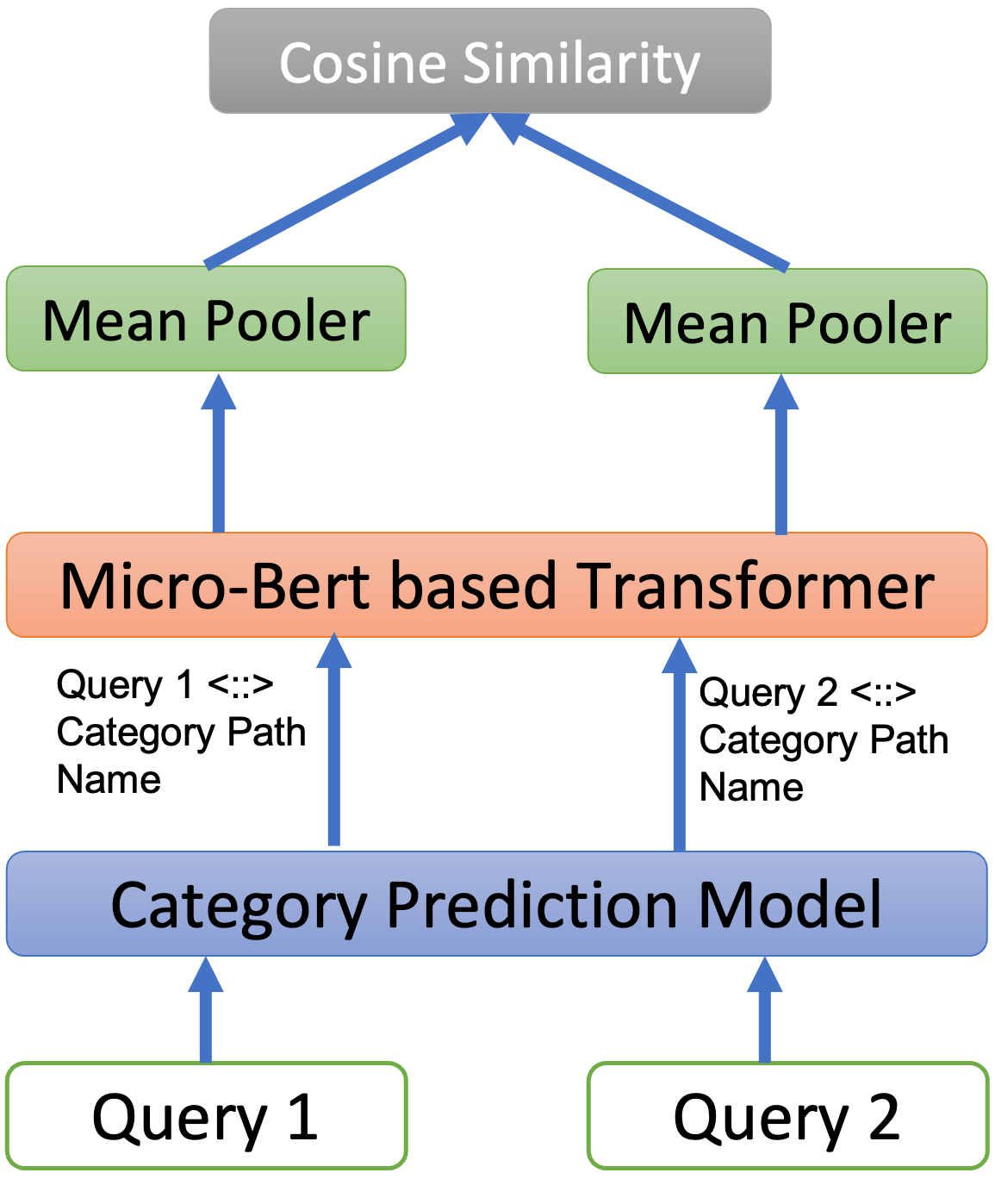}
\centering
\caption{Model for Online Query Similarity}
\label{fig:figure model}
\end{figure}

We use a two tower model with a contrastive learning loss function to train a query representation, which can be used to find similar queries. Figure \ref{fig:figure model} shows the  model architecture. We use a micro-BERT model pre-trained on e-Commerce titles to obtain the embedding for a query by concatenating it with the output of the query-category classifier, and then aggregating the subword embeddings vectors using mean. We measure the similarity between two queries using cosine between their embeddings to fine tune the micro-BERT model.
\section{Online Nearest Neighbor Approach}
As discussed earlier, we make offline use of a nearest neighbor database to sample pairs of similar queries.  We also use this nearest neighbor database to support online processing of unseen queries.

Specifically, we populate a nearest-neighbor database with the vectors of known queries that cover the head and torso of the query distribution. We start with the most frequent 10 million queries from relatively recent behavior (the past year) and then filter this set to enforce some quality criteria. We remove misspellings and queries returning too few results. We also remove near-duplicate queries that are too similar to better queries -- where "better" is a function of query frequency and result set size.

When the searcher enters a query -- especially a query that would otherwise return no or few results -- we can look it up in this nearest-neighbor database and return the known query or queries most similar to it. Then, if the similarity score is high enough (e.g., >= 0.98), the search engine can use it to increase recall or to leverage signals from the historical behavior associated with these known queries.

\section{Examples}

Here are examples of query pairs from the Amazon Shopping Queries that have high behavioral similarity despite low surface similarity:

\begin{table}[h]
\tiny
\begin{tabular}{|p{2.8cm}|p{1.8cm}|p{1.7cm}|}
\hline
\textbf{Query 1} & \textbf{Query 2} & \textbf{Cosine Similarity} \\
\hline
hdmi to galaxy s8 & s9 hdmi & 0.9993 \\
\hline
movie money & prop money & 0.9995 \\
\hline
cassette adapter for iphone & tape to aux & 0.9993 \\
\hline
\end{tabular}
\label{tab:my_label}
\end{table}

Conversely, here are some examples of query pairs that have low behavioral similarity despite high surface similarity:
\begin{table}[h]
\tiny
\begin{tabular}{|p{2.8cm}|p{1.8cm}|p{1.7cm}|}
\hline
\textbf{Query 1} & \textbf{Query 2} & \textbf{Cosine Similarity} \\
\hline
alice in wonderland & alice in wonderland bow & 0.790806 \\
\hline
mud & pink mud & 0.676611 \\
\hline
the rent collector book & the bone collector book & 0.768282 \\
\hline
\end{tabular}
\label{tab:my_label}
\end{table}

These examples, while anecdotal, illustrate how the proposed approach for recognizing similar or equivalent search intent is more flexible and robust than relying directly on the surface query tokens.

\section{Experiments}
Unfortunately, there are no benchmarks available to directly evaluate query similarity for e-Commerce search. To evaluate the proposed query similarity approach using public data, we make indirect use of the Amazon Shopping Queries data set \cite{reddy2022shopping}, augmented with structured data crawled from the Amazon site
 \cite{esci-s}.
 
In addition, we evaluate our approach using a proprietary collection of eBay queries that have a a fraction of co-engaged items.

For both the public and proprietary datasets, we evaluate the performance of our model by comparing the category distributions of clicks for the queries -- information that we do not make available to the models. Specifically, we compare the distributions using Pearson correlation.

All experiments were performed on a server running Ubuntu 20.04 with 128GB of RAM and a single NVIDIA v100 GPU.

As shown in the table below, our approach outperforms all-MiniLM-L12-v2, a sentence transformers model, on our proxy for query similarity, with a Pearson correlation of 0.85 vs. 0.68.
\begin{table}[h]
\begin{tabular}{|p{2.8cm}p{1.8cm}p{1.8cm}|}
\hline
\textbf{Model} & \textbf{Dataset} & \textbf{Pearson's r} \\ 
\hline
query-sim-ecom & eBay internal &  0.87 \\
\hline
query-sim-ecom & esci query-query &  0.85 \\
\hline
all-MiniLM-L12-v2 & esci query-query &  0.68 \\
\hline
\end{tabular}
\label{tab:my_label}
\end{table}
\section{Conclusion}

The proposed approach demonstrates that we can measure e-Commerce query similarity and recognize equivalent queries by aggregating historical searcher behavior for frequent queries and training a model that generalizes to unseen queries. While the approach presented relied on vectors generated from product titles, it can be generalized to richer representations, such as multimodal product representations that combine text, structured data, and images. We also hope to see the development of query similarity benchmarks that make it possible to directly compare solutions for this important e-commerce problem.

\bibliography{sing}
\bibliographystyle{acl_natbib}

\end{document}